\documentclass[sigplan,10pt]{acmart}
\renewcommand\footnotetextcopyrightpermission[1]{} 
\acmConference{}

\startPage{1}

\setcopyright{none}

\usepackage{booktabs}   
\usepackage{subcaption} 
\input{packages}

\begin{document}

\title{Probabilistic Programming with CuPPL} 


\author{Alexander Collins}
\affiliation{
  \institution{NVIDIA}
}
\email{acollins@nvidia.com}

\author{Vinod Grover}
\affiliation{
  \institution{NVIDIA}
}
\email{vgrover@nvidia.com}

\begin{abstract}

Probabilistic Programming Languages (PPLs) are a powerful tool in machine learning, allowing highly
expressive generative models to be expressed succinctly. They couple complex inference algorithms,
implemented by the language, with an expressive modelling language that allows a user to
implement any computable function as the generative model.

Such languages are usually implemented on top of existing high level programming languages and do
not make use of hardware accelerators. PPLs that do make use of accelerators exist, but restrict the
expressivity of the language in order to do so.

In this paper, we present a language and toolchain that generates highly efficient code for both
CPUs and GPUs. The language is functional in style, and the toolchain is built on top of LLVM. Our
implementation uses delimited continuations on CPU to perform inference, and custom CUDA codes on
GPU.

We obtain significant speed ups across a suite of PPL workloads, compared to other state of the art
approaches on CPU. Furthermore, our compiler can also generate efficient code that runs on CUDA
GPUs.

\end{abstract}


\maketitle

\thispagestyle{fancy}

\section{Introduction}
\sectionlabel{introduction}

Probabilistic Programming Languages (PPLs) are a powerful machine learning tool, allowing highly
expressive generative models to be expressed succinctly, and complex inference to be applied to
them. They typically provide a range of built-in distributions from which samples can be drawn and
scored, along with a range of inference algorithms.

\Nova is a functional programming language, with all of the standard control flow constructs
including let expressions, conditional expressions, recursive functions, algebraic datatypes and
case statements. It also includes a \code{vector} data type and a suite of built-in vector operators
such as \code{map}, \code{reduce} and \code{repeat}.
\Nova provides two basic probabilistic primitives: \code{sample} and \code{factor}. These can be
used within a generative model to sample from and score a suite of built-in distributions and user
defined distribution objects.
These primitives are implemented in using delimited continuations and the operators \code{shift} and
\code{reset} \cite{Felleisen1988}.
\Nova then provides inference methods that can be applied to a model function in order to generate
the posterior distribution from the model.

\subsection{PPL Example}

An example of a generative model is given in \figureref{ppl-example2}, which
demonstrates the diverse control flow that \nova allows within the model function, and which gives
PPLs their expressive power.

This model finds the $n$-degree polynomial that best fits some given data. The degree of the
polynomial is sampled from a uniform prior, and its coefficients from gaussian priors. The
polynomial is then conditioned on a measure of its ``distance'' from the data. The model returns the
polynomial, as a vector of coefficients. Running inference on this model produces an empirical
distribution over polynomials, and the mode of this distribution is the ``line of best fit'' for the
data.

\begin{figure}
  \begin{novacode}
    model <- function() {
      n <- sample(uniform-discrete(2,5));
      line <- repeat(
        function(i) { sample(normal(0,10)) },
        n);
      factor(-distance(line, data));
      line
    }
  \end{novacode}
  \caption{
    An example program that demonstrates the use of universal language features in a model
    function. The implementation of \code{distance} is omitted for brevity.
  }
  \figurelabel{ppl-example2}
\end{figure}


\subsection{Delimited Continuations}

Whenever the model function calls \code{sample} or \code{factor}, or returns a value, the inference
algorithm needs to make decisions about its execution -- more specifically the probability of a
trace of execution through the model needs to be tracked and samples need to be generated.

We use delimited continuations (the \code{shift} and \code{reset} operators) to allow the inference
algorithm to manipulate the execution of the model function and generate samples from its posterior
distribution.

\code{shift(k, e)} captures the current state of the program up to the last enclosing
\code{reset(.)}, creates a function from this that can be called via $k$, and evaluates $e$. $e$ may
call $k$ as few or as many times as it likes.

Using \code{shift} whenever \code{sample} or \code{factor} are called captures the current state of
execution of the model function (up to the start of the execution of the model function) and allows
the inference algorithm to decide what to execute next.

\sectionref{language} describes our implementation of \code{sample} and \code{factor} in more detail.

%
%

\subsection{Contributions}

In this paper, we make the following contributions:

\begin{itemize}
  \item
    We describe a probabilistic programming language that is functional in nature, and supports
    delimited continuations using the shift and reset operators described in \cite{asai07}. The
    language uses answer type polymorphism in its type system to statically type check the code
    enabling statically typed LLVM IR to be generated from it.
  \item
    A novel implementation strategy for delimited continuations in LLVM IR that enables inference to
    be implemented in LLVM IR in a similar style to that described in \cite{Goodman2013}.
  \item
    The code generated by our compiler performs significantly better than both Anglican
    \cite{tolpin2016design} and webppl \cite{dippl} over a range of benchmark programs.
  \item
    Moreover, the code generated by our compiler also scales well to large numbers of samples.
\end{itemize}

\subsection{Structure}

The rest of the paper is structured as follows. In \sectionref{relatedwork} we discuss related work,
in \sectionref{language} we give an overview of the features and type system \nova, and in
\sectionref{codegen} we discuss the design and implementation of the code generator. This is
followed in \sectionref{evaluation} by a performance analysis of our language across a range of
benchmarks, comparing against both webppl \cite{dippl} and Anglican
\cite{tolpin2016design}. \sectionref{conclusion} concludes the paper.

\section{Related Work}
\sectionlabel{relatedwork}

The probabilistic programming languages most similar to our own are Anglican \cite{tolpin2016design}
and webppl \cite{dippl}, however both of these languages are embedded in other programming
languages. The semantics of our language also borrows heavily from \cite{Goodman2013} which
describes many of the principles behind implementing probabilistic languages.

Anglican \cite{tolpin2016design} is embedded in Clojure \cite{clojure}, a dialect of Lisp that works
in the Java Virtual Machine. In is built on top of continuations and trampolines in order to capture
control and pass it to the inference algorithm.

webppl \cite{dippl} is written in Javascript, on top of the NodeJS ecosystem \cite{nodejs}. It uses
coroutines to capture control and pass it to the inference algorithm.

Similarly to Anglican and webppl, Kiselyov et al \cite{Kiselyov2009} describe a domain specific
probabilistic language that is embedded in OCaml. They also use the delimited continuation approach
to explore programs as models.

Edward \cite{tran2016edward,tran2017deep} is a probabilistic modelling library for TensorFlow
\cite{tensorflow2015-whitepaper}. This is not a true PPL as it does not allow arbitrary control
flow, and is limited by the graph structures that are representable in TensorFlow.

Compiler based techniques to producing PPL code include \cite{pmlr-v15-wingate11a}. This approach
translates a PPL problem into MCMC code, rather than compiling the model function in the language
itself. TerpreT \cite{TerpreT} is a domain specific probabilistic language for program induction,
generating programs from a PPL model.

Earlier approaches to probabilitic programming and modelling include BUGS \cite{BUGS}, which is
restricted to Bayesian modelling and cannot handle the complex control structure that true PPLs can.

Stan \cite{Stan}, similarly to BUGS, is restricted to Bayesian modelling. Bindings exist for
embedding Stan in either R or Python.
Other early, and similar, approaches include Church \cite{Church}.

Borgstrom et al \cite{Borgstrom2016} present a rigorous lambda calculus for probabilistic
programming, in the style of \nova, Anglican, webbpl and Church. They use this to prove that they
perform MCMC inference on program traces correctly.

\section{The Language}
\sectionlabel{language}

\Nova is a statically-typed functional language, and its design is centered around vectors and data
parallel operations. It is also designed to allow high-level language transformations such as
deforestation of vector operations, closure conversion and other high level optimization passes.

The rest of this section is structured as follows.
\sectionref{language-features} describes the salient features of \nova.
\sectionref{delimited-continuations} describes how delimited
continuations are expressed, \sectionref{sample-and-factor} details how these are used to implement
\code{sample} and \code{factor}, and finally \sectionref{type-system} details how the polymorphic
type system treats delimited continuations using answer types.

\subsection{Language Features}
\sectionlabel{language-features}

\Nova includes the usual control flow constructs you would expect from a functional language
including lambda expressions, let-expressions, conditionals, case statements and recursion.
The language also includes first class vector data types of arbitrary dimensions, and a suite of
parallel operators that work over them, including:

\begin{itemize}
  \item
    \code{map} applies a function to every element of the vector
  \item
    \code{repeat} generates a vector by running a function a given number of times
  \item
    \code{reduce} performs a parallel reduction over a vector
  \item
    \code{filter} constructs a new vector that includes only those elements from the input vector
    for which a given predicate returns true
\end{itemize}

\subsubsection{Type Generalization and Specialization}

\Nova allows type generalization and specialization, in a similar manner to System F. 
Polymorphic types can be defined as follows:
\begin{novacode}
type List 'a :
  (+ (Nil : unit)
     (Cons : ('a, (List 'a))));
\end{novacode}

This example demonstrates both type generalization and type specialization. List is the type
constructor:
\begin{novacode}
forall 'a .
  (+ (Nil : unit)
     (Cons : ('a, (List 'a))))
\end{novacode}

This type can be specialized to store a list of integers by using a type application:
\begin{novacode}
(List int)
\end{novacode}

This produces the concrete type:
\begin{novacode}
(+ (Nil : unit)
   (Cons : (int, (List int))))
\end{novacode}

We impose a few restrictions on the use of generalized types, \code{forall} types and type
applications. Firstly, general types can only be constructed at the start of a module, as the type
constructors need to be unique within a module. Secondly, after type checking a program, all types
must be specialized, or turned into concrete types. This is because the LLVM IR target environment
requires all types to be concrete. If an expression is discovered whose type is not specialized to a
concrete type (such as partial application of a parallel operator), the compiler complains that it
could not statically determine the concrete type of the expression.

\subsection{Distribution Values}

The type system includes a distribution type, denoted \code{\~t} which is a distribution over values
of type \code{t}. It also includes a suite of built-in distributions, including gaussian, bernoulli,
poisson and many others. Built-in functions to perform operations on distribution values are also
provided, such as \code{sample*} which produces a sample from a distribution, \code{dist-score}
which scores a sample from a distribution and \code{dist-var} which returns the variance of a
distribution.

\subsection{Delimited Continuations}
\sectionlabel{delimited-continuations}

\Nova allows delimited continuations through the use of the \code{shift} and \code{reset}
constructs \cite{Felleisen1988, asai07}.

Informally, the meaning of \code{shift} is to capture the current execution state up to the last
\code{reset}, wrap it in a function, and bind a name to that function.

For example, consider the following:
\begin{novacode}
reset(1 + shift(k, k(2)))
\end{novacode}

When \code{shift} is encountered, it captures the current program as a function named $k$. In this
case $k$ is:
\begin{novacode}
function (x) { 1 + x }
\end{novacode}
It then executes the body of the shift expression, which executes \code{k(2)}, so the program
returns 3.

Reset limits the scope to which shift will capture the state of the program. For example, consider
the following:
\begin{novacode}
1 + reset(3 + shift(k, 1))
\end{novacode}
When \code{shift} is encountered, it captures the current program as a function $k$ up to the last
reset. In this case $k$ is:
\begin{novacode}
function (x) { 3 + x }
\end{novacode}
It then executes the body of the shift expression, which simply returns \code{1} and does not
execute $k$, so the program returns 2 as its result.

\subsection{Inference, sample and Factor}
\sectionlabel{sample-and-factor}

The two basic PPL primitives \code{sample} and \code{factor} are expressed in terms of delimited
continuations. \figureref{sample-factor-observe-impl} gives the definition of \code{sample} and
\code{factor} (and the \code{observe} helper function) in terms of \code{shift} and \code{reset}.

\begin{figure}
  \begin{novacode}
  sample <- function (dist : ~a) : a {
    shift (k : a => b) {
      x <- sample-impl(dist, k);
      (x[0])(x[1])
    }
  }
  \end{novacode}
  \begin{novacode}
  factor = function (log-p : a) : unit {
    shift k {
      f <- factor-impl(log-p, k);
      f()
    }
  }
  \end{novacode}
  \begin{novacode}
  observe = function (dist, x) : unit {
    factor(dist-score(dist, x))
  }
  \end{novacode}
  \caption{
    Implementation of \code{sample}, \code{factor} and \code{observe} using delimited continuations
  }
  \figurelabel{sample-factor-observe-impl
  }
  \figurelabel{sample-factor-observe-impl}
\end{figure}

\code{sample} uses the \code{shift} operator to capture the current execution state of the model
function as a function $k$. When inference starts, \code{reset} is called to limit the scope of this
capture up to the start of the execution of the model function. \code{sample} then calls
\code{sample-impl} with the distribution object that was sampled from and $k$. The implementation of
\code{sample-impl} depends on the choice of inference algorithm, however it always returns a tuple
of type \code{(a => b, a)}. This function represents what the program should do next: the first
element of the tuple is a continuation, with the same type as $k$, and the second element of the
tuple is a sample from the distribution, of type \code{a}.

\code{factor} also uses \code{shift} operator to capture the current execution state of the model
function as a function $k$. It then calls \code{factor-impl} with the log probability with which to
update the program trace. The implementation of \code{sample-impl} depends on the choice of
inference algorithm, however it always returns a function of type \code{unit => unit}. This is a
function denoting what the program should do next, and has the same type as $k$.

\code{observe} is a wrapper function around \code{factor} that makes altering the log probability of
the execution trace based on observing data from a distribution simpler. It calls \code{factor} with
the score of the given sample \code{x} from the given distribution \code{dist}.

\Nova provides two inference methods that implement \code{sample-impl} and \code{factor-impl}. They
are invoked via built-in functions called \code{importance} (for running importance sampling) and
\code{mcmc} (for running Monte Carlo Marov Chain based inference. More details on the implementation
of these functions can be found in \sectionref{codegen}.

\subsection{Type System}
\sectionlabel{type-system}

\Nova uses Hindley-Milner type inference, augmented with polymorphic answer types
\cite{asai07}. \figureref{type-rules} gives a subset of the type rules. Te following notation is
used:

\begin{itemize}
\item
  Types are denoted $\tau$.
\item
  Type schemes are denoted $\sigma$. They have the syntax
  $\sigma = \tau \mid \forall A . \tau$ where $A$ is a set of type variables.
\item
  $\tau \prec \sigma$ denotes that type $\tau$ is an instance of type scheme
  $\sigma$
\item
  Answer types are denoted $\alpha$, $\beta$, $\gamma$. Answer types can be any type
  $\tau$ or the empty type, denoted $\bot$
\item
  Expressions are denoted $e$. 
\item
  Constant values are denoted $c$. 
\item
  Variable identifiers are denoted $x$. 
\item
  Type environments are denoted $\Gamma$. These map variable identifiers $x$ to
  type schemes $\sigma$. The notation $\Gamma\{x : \sigma\}$ denotes mapping
  $\Gamma$ updated with identifier $x$ mapped to type scheme $\sigma$.
  $\Gamma_0$ denotes an empty type environment, that does not map any
  identifiers to type schemes.
\item
  There are two forms of type judgements:
  \begin{itemize}
  \item
    \emph{Pure} type judgments are written $\Gamma \vdash_p e : \tau$. This means that
    expression $e$ has type $\tau$, given type environment $\Gamma$.
  \item
    \emph{Impure} type judgments are written $\Gamma,\alpha \vdash e : \tau,\beta$.
    This means that expression $e$ has type $\tau$, given type environment
    $\Gamma$ and the execution of $e$ changes the answer type from $\alpha$ to
    $\beta$.
  \end{itemize}
\end{itemize}

\begin{figure}
  \begin{typerules}
  \typerule{lambda}
    {\Gamma\{x : \tau_1\}, \alpha_1 \vdash e : \tau_2, \alpha_2}
    {\Gamma \vdash_p \term{function [}\alpha_1 \alpha_2\term{] (} x \term{:} \tau_1 \term{):} \tau_2 \term{ \{ } e \term{ \}} : \tau_1 \term{/} \alpha_1 \rightarrow \tau_2 \term{/} \alpha_2} \\
  \typerule{apply}
    {\Gamma,\alpha_1 \vdash e_1 : \tau_2 \term{/} \alpha \rightarrow \tau \term{/} \alpha_2, \beta \and
     \Gamma,\alpha_2 \vdash e_2: \tau_2,\alpha_1}
    {\Gamma,\alpha \vdash e_1 \term{(} e_2 \term{)} : \tau,\beta} \\
  \typerule{reset}
    {\Gamma,\alpha \vdash e : \alpha,\tau}
    {\Gamma \vdash_p \term{reset \{ } e \term{ \}} : \tau} \\
  \typerule{shift}
    {\Gamma\{x : \forall t.( \tau \term{/} t \rightarrow \alpha \term{/} t )\}, \tau_2 \vdash e : \tau_2,\beta }
    {\Gamma,\alpha \vdash \term{shift } (x : \tau_1) : \tau_2 \term{ \{ } e \term{ \}} : \tau,\beta} \\
  \typerule{expr}
    {\Gamma \vdash_p e : \tau}
    {\Gamma,\alpha \vdash e : \tau,\alpha}
  \end{typerules}
  \caption{
    Subset of the type rules for \nova.
  }
  \figurelabel{type-rules}
\end{figure}

The \rulename{expr} rule allows an expression with the same input and output answer types to be type
checked as a pure expression.

\subsection{Inference Methods}
\sectionlabel{infer-impl}

\Nova supports three inference methods.

\subsubsection{Enumeration}

Enumeration performs precise inference on the model function, by executing all possible paths
through the function.  The probabilistic branching points are when the model calls
\code{sample}. Whenever \code{sample} is called, the inference algorithm executes the model function
from that point for all values in the support of the distribution. This only works for distributions
with finite support, and returns a runtime error if a continuous distribution is encountered. This
performs a breadth first traversal through the model function, and the depth of this traversal can
be limited to a maximum depth if desired.

\subsubsection{Importance Sampling}

This performs importance sampling \cite{StochasticSimulation} on the model function. This is done by
repeatedly running the model function. The calls to \code{sample} and \code{factor} made by each
execution prescribe a log probability the value returned by the model function. The model function
is repeated up to a chosen number of times, and the posterior distribution is computed by
normalizing the log probabilities of the returned values.

\subsubsection{Lightweight Metropolis-Hastings}

This inference method \cite{pmlr-v15-wingate11a} uses the Metrolpolis-Hastings algorithm to generate
samples from the posterior distribution of the model function. First, the model function is run once
to memorize the return values of any calls to \code{sample}. To generate a new sample, One of the
memorized return values is chosen uniformly randomly, re-sampled, and the model function is
re-executed from this point. Subsequent calls to sample reuse memorized values where possible. This
may not be possible if the modification affects the control flow of the program, in which case new
values are sampled and memorized. When the model function completes, the Metropolis-hasting
acceptance criteria is used to either accept or reject the sample.

\section{Code Generation}
\sectionlabel{codegen}

Our compiler parses the code, performs language-specific optimization passes on the AST, and then
generates code for either CPU or GPU using a code generator built on top of the LLVM toolchain
\cite{LLVM:CGO04}.

The rest of this section is structured as follows. \sectionref{codegen-language-features} describes
how various language features are compiled to LLVM IR, including closures and distribution
objects. \sectionref{codegen-stack} describes the implementation of the call stack and how different
language constructs interact with it. \sectionref{codegen-delimcont} describes how delimited
continuations are implemented for CPU. Finally \sectionref{codegen-inference} describes the code
generation for the inference methods on CPU, using delimited continuations, and GPU, using custom
generated NVVM IR.

\subsection{Language Features}
\sectionlabel{codegen-language-features}

\subsection{Call Stack}
\sectionlabel{codegen-stack}

The generated LLVM IR does not use the \llvmir{call} operator. Instead it uses a combination of
\llvmir{indirectbr} and a custom ``managed stack''. This allows for more complex stack operations
necessary for implementing delimited continuations as discussed in \sectionref{codegen-delimcont}.

The managed stack is stored in region of heap allocated memory, allocated when the program starts. A
stack pointer is also stored. LLVM IR is generated for the operations that modify the stack:

\begin{itemize}
  \item
    Push and pop work just like a conventional stack.
  \item
    The stack save operation copies a chosen number of bytes from the top of the stack into a newly
    allocated region on the heap, and returns a pointer to it, and then pops this region off of the
    stack. This operation is used by delimited continuations to capture the current state of the
    program up to the last reset, as part of the implementation of \code{shift} discussed in
    \sectionref{codegen-delimcont}.
  \item
    The stack restore operation pushes a previously saved stack region onto the top of the
    stack. This operation is used to restore the state of the stack when calling a continuation that
    was captured by the \code{shift} operator, as detailed in \sectionref{codegen-delimcont}.
\end{itemize}

\subsubsection{Closures}

The LLVM IR code for a closure begins with a labelled basic block. A closure is called using an
\llvmir{indirectbr} to the address of this block (obtained using \llvmir{blockaddress}). This label
is annotated with all possible predecessor blocks, as required by LLVM IR. These are statically
known, but may be a superset of all possibly call sites at runtime.

A closure value is represented by an LLVM IR struct of type \llvmir{\{i64,i64\}}, containing a
pointer to its code and and some memory storing its environment. The environment of a closure is
represented by an LLVM IR struct containing the values of its captured variables. If the environment
is empty, this struct has type \llvmir{\{\}}. The instructions in the entry block of the closure
first unpack this environment into registers, before continuing with the execution of the closure.

When calling a closure, the LLVM IR \llvmir{call} instruction is not used, so that more complex
stack operations can be handled. This requires the code generator to generate code for entry and
return from the call. The code generation strategy is summarised by the pseudocode in
\figureref{codegen-closures}.

\begin{figure}
  \begin{pseudocodenumbered}
  generateClosureCall(entryPtr, envPtr, args) {
    retBlk = llvm::BasicBlock::Create();
    retAddr = llvm::BlockAddress::get(retBlk);
    for (var in symTab)
      generateStackPush(var);
    generatePush(retAddr);
    generatePush(envPtr);
    for (arg in args)
      generateStackPush(arg);
    builder.CreateIndirectBr(entryPtr);
    builder.SetInsertPoint(retBlk);
    result = generatePop();
    for (var in reverse(symTab))
      symTab.update(var, generateStackPop());
    return result;
  }
  \end{pseudocodenumbered}
  \begin{pseudocodenumbered}
  generateClosureEntry(argVars, freeVars) {
    for (var in reverse(argVars))
      symTab.update(var, generateStackPop());
    envPtr = generateStackPop();
    for (var in freeVars)
      symTab.update(var, unpack(envPtr, var));
  }
  \end{pseudocodenumbered}
  \begin{pseudocodenumbered}
  generateClosureExit(result) {
    retAddr = generateStackPop();
    generateStackPush(result);
    builder.CreateIndirectBr(returnAddress);
  }
  \end{pseudocodenumbered}
  \caption{Code generation strategy for closures}
  \figurelabel{codegen-closures}
\end{figure}

\subsubsection{Distribution Values}

Built-in distributions are represented by a struct of type \llvmir{\{int32, int64, int64,
  int64\}}. The first element is a tag determining the type of the distribution, and the subsequent
elements contain distribution specific data. These three values provide sufficient space to store
parameters for any of the built-in distributions. This fixed sized encoding of distribution values
allows them to be treated as plain values in LLVM IR, avoiding the need for heap allocation. This
makes them suitable for use on a GPU where heap allocations are expensive.

\subsubsection{Built-in Functions and Runtime Library}

\Nova includes a runtime library which implements the built-in functions that are not directly
available in the LLVM IR instruction set. For example, sampling and scoring distributions is
implemented in the runtime library. Separate implementations of this library are provided for CPU
and CUDA GPUs.

\subsubsection{Garbage Collection}

On CPU, where it is needed, the Boehm-Demers-Weiser Garbage Collector \cite{bdwgc} is used.  On GPU,
garbage collection is not feasible. Static shape analysis is used to determine a fixed upper bound
for the size of any vectors used so that they can be stack allocated. If the size of a vector cannot
be statically determined the code generator will fall back to making a heap allocation. Depending on
the structure of the program, this heap allocation fallback may exhaust the heap space available on
the GPU, in which case the program will have to be reworked.

\subsection{Delimited Continuations}
\sectionlabel{codegen-delimcont}

The \code{reset} operator marks the current state of the program. It does this by pushing the value
of the stack pointer onto the top of a separate ``reset stack''. When the reset operator exits, the
value is popped off of the reset stack.

The \code{shift} operator captures the current program state as closure value that can be called at
some later point. It creates a closure value containing a pointer to the current instruction and the
contents of the stack up to the last call to \code{reset}. The address of the current instruction is
obtained by starting a new basic block before that instruction and obtaining its address using
\llvmir{blockaddress}. A copy of the stack is obtained using the stack copy operation, to save the
region from the current stack pointer up to the stack pointer stored on the top of the reset stack.

The environment for the captured continuation differs from that of a regular closure. Instead of a
struct containing the values of any captured variables, it is a struct containing the saved stack
region. It has type \llvmir{\{int32, int8*\}} -- storing the size of the captured stack and a
pointer to it.

When a captured continuation is called the same steps are taken as calling a regular
closure. However, instead of unpacking the environment structure into local variables, the saved
stack from the environment structure and is pushed onto the top of the call stack. This restores the
stack (up to the last reset) to the same state it was in when the continuation was captured.

Pseudocode for the implementation of this code generation is shown in \figureref{codegen-delimcont}.
\pseudo{generateResetEntry} generates code for the entry into a \code{reset} expression, and returns
the exit block so that the caller can pass it to \pseudo{generateResetExit} later.
\pseudo{generateResetExit} is called, with the exit block and the return value of the
sub-expression, after code has been generated for the sub-expression in the \code{reset}.
\pseudo{generateShiftEntry} and \pseudo{generateShiftExit} generate code for the entry and exit from
a \code{shift} expression.

\begin{figure}
  \begin{pseudocodenumbered}
  generateResetEntry() {
    for (var in symTab)
      generateStackPush(var);
    exitBlock = llvm::BasicBlock::Create();
    exitAddress = llvm::BlockAddress::get(
      exitBlock);
    generateStackPush(exitAddress);
    stackPtr = generateGetStackPtr();
    generateResetStackPush(stackPtr);
    return exitBlock;
  }
  \end{pseudocodenumbered}
  \begin{pseudocodenumbered}
  generateResetExit(exitBlock, value) {
    returnAddress = generateStackPop();
    generateStackPush(value);
    builder.CreateIndirectBr(returnAddress);
    builder.SetInsertPoint(exitBlock);
    result = generateStackPop();
    generateResetStackPop();
    for (var in reverse(symTab))
      symtab.update(var, generateStackPop());
    return result;
  }
  \end{pseudocodenumbered}
  \begin{pseudocodenumbered}
  generateShiftEntry() {
    for (var in symTab)
      generateStackPush(var);
    kBlock = llvm::BasicBlock::Create();
    kAddress = llvm::BlockAddress::Get(kBlock);
    startPtr = generateResetStackPeek();
    capturedStack = generateStackSave(startPtr);
    k = builder.CreateStruct(
      kAddress, capturedStack);
    return builder.CreateStruct(
      k, kBlock);
  }
  \end{pseudocodenumbered}
  \begin{pseudocodenumbered}
  generateShiftExit(k, kBlock, value) {
    returnAddress = generateStackPop();
    generateStackPush(value);
    builder.CreateIndirectBr(returnAddress);
    builder.SetInsertPoint(kBlock);
    argument = generateStackPop();
    capturedStack = generateStackPop();
    generateStackRestore(capturedStack);
    for (var in reverse(symTab))
      symTab.upate(var, generateStackPop());
    return argument;
  }
  \end{pseudocodenumbered}
  \caption{
    Pseudocode for code generation of delimited continuation operators \code{shift} and \code{reset}
  }
  \figurelabel{codegen-delimcont}
\end{figure}

\subsection{Inference}
\sectionlabel{codegen-inference}

Code is generated for the model function, which uses \code{shift} expressions to capture the model
functions continuation and pass it to \code{sample-impl} and \code{factor-impl}, as described in
\sectionref{sample-and-factor}, and in the previous section.

Code is generated for the inference methods \code{importance} and \code{mcmc}, first by converting
them to calls to \code{infer-impl}, as described in \sectionref{infer-impl}, and then generating the
code for the resulting abstract syntax tree.

\section{Evaluation}
\label{sec:evaluation}

This section explores the performance of \nova.
In \sectionref{ppl-performance} we compare \nova against webppl \cite{dippl} and Anglican
\cite{tolpin2016design} across a range of benchmarks, including probabilistic models and
micro-benchmarks.
In \sectionref{delimcont-performance} we compare the performance of our implementation of delimited
continuations against Racket \cite{plt-tr1}.

\subsection{Performance vs Other PPLs}
\sectionlabel{ppl-performance}

The experiments were run on an i7-5820K CPU with 16 GB of main memory and an NVIDIA GTX 710. The
wall clock execution time of the entire program is measured for 10 repeats. Error bars in the plots
show the standard deviation of the measured data.

The benchmark programs are summarized in \tableref{benchmarks}. Each benchmark was implemented in
Anglican \cite{tolpin2016design}, webppl \cite{dippl} and \nova. The \nova versions of the benchmark
code were compiled for both CPU and GPU and run separately. Anglican does not support exact
inference, therefore the \verb|enumerate_geometric| benchmark is omitted for this language.

For models using importance sampling, each language is configured to generate 2 million samples, For
MCMC inference, each language is configured to generate 100 thousand samples and for exact
inference, 10 thousand maximum executions are used. These values were chosen to provide execution
times of the order of a few seconds across most benchmarks.

\begin{figure}
  \footnotesize
  \rowcolors{2}{white}{gray!25}
  \begin{tabular}{lp{5cm}}
    \verb|biasedcoin| & Importance sampling to compute expected value of a biased coin,
                        conditioned on observed coin flips \\
    \verb|customdist| & Importance sampling from a custom distribution
                        (sum of two gaussians) \\
    \verb|linear_regression| & Linear regression using MCMC \\
    \verb|logistic_regression| & Logistic regression using MCMC \\
    \verb|binomial| & Importance sampling to generate a binomial distribution \\
    \verb|sevenscientists| & Importance sampling for the seven scientists problem \\
    \verb|linefitting| & n-degree polynomial line fitting using importance sampling \\
    \verb|enumerate_geometric| & Exact inference to generate a geometric distribution \\
  \end{tabular}
  \caption{Benchmark programs}
  \tablelabel{benchmarks}
\end{figure}

The results are shown in \figureref{results}. This plot shows that \nova is significantly more
performant that both webppl and Anglican across all but one of the benchmarks.
webppl fails to run the line fitting benchmark -- the program was terminated
after 10 minutes without producing a result.

The GPU implementation of \nova improves performance compared to the CPU implementation, for most of
the importance sampling benchmarks. GPU inference does not yet support MCMC or exact inference. The
customdist benchmark performs similarly on CPU and GPU, most likely due to the branching nature of
the benchmark. The more complex line fitting benchmark also shows less of a performance improvement
on GPU compared to CPU. This is also likely due to the branching nature of the program, which is
dependent on an integer sampled from the uniform discrete distribution.

\begin{figure}
  \includegraphics[width=0.98\columnwidth]{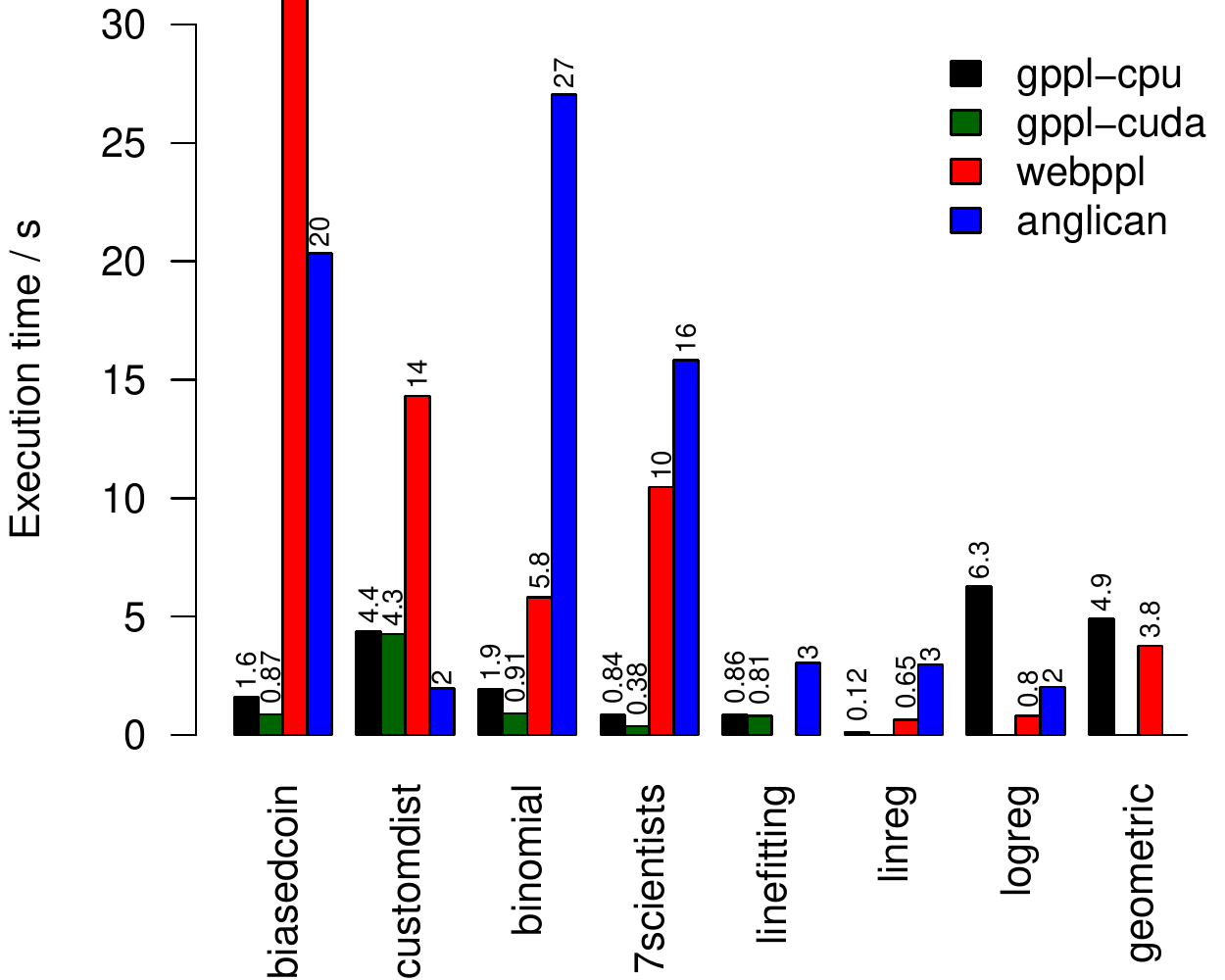}
  \caption{
    Execution time across all benchmarks and languages.
    Error bars show one standard deviation for 10 repeated measurements.
    Where data is omitted, the benchmark is either not supported by that language or takes more than
    10 minutes to run.
  }
  \figurelabel{results}
\end{figure}

The scalability for two of the benchmarks is shown in \figureref{scaling-ppl}. \verb|linefitting| is
the more complex of the importance sampling benchmarks, and \verb|linear_regression| uses
Lightweight MH inference. These results show that each language scales similarly, however the slope
for \nova is reduced.

\begin{figure}
  \includegraphics[width=0.98\columnwidth]{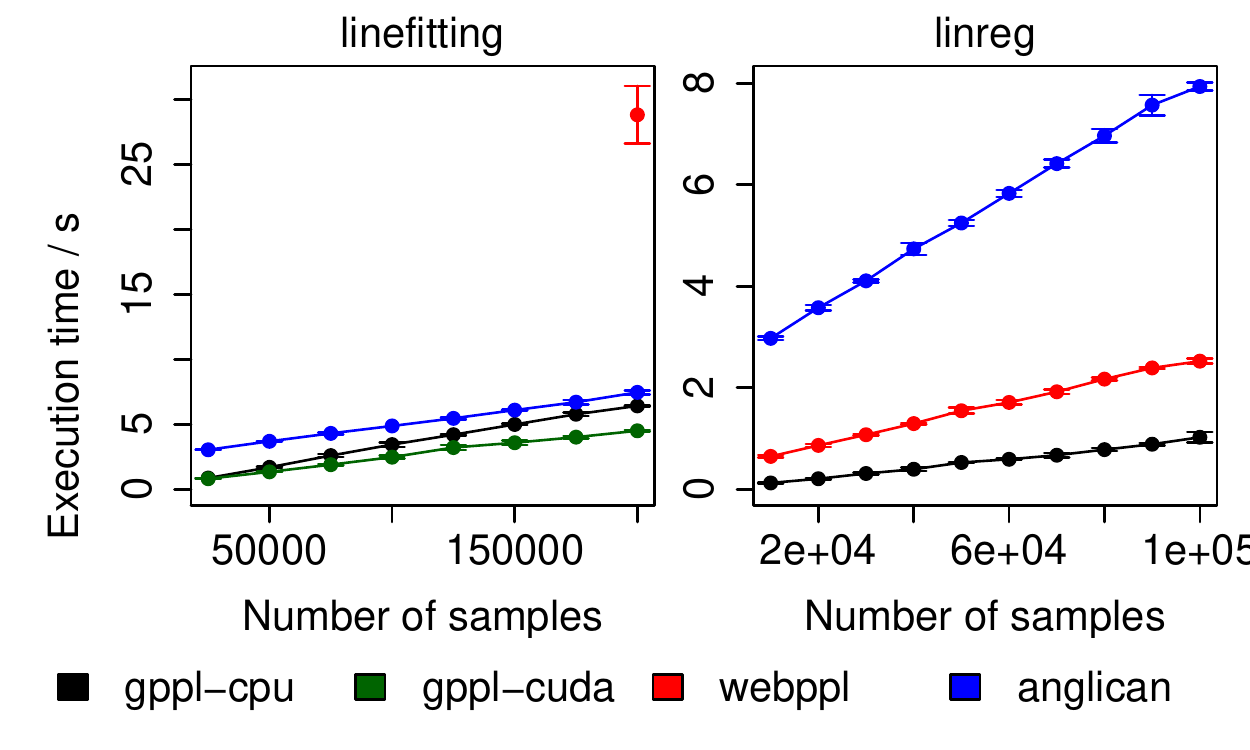}
  \caption{
    Performance two benchmarks using different inference methods.
    Error bars show one standard deviation for 10 repeated measurements.
  }
  \figurelabel{scaling-ppl}
\end{figure}

\subsection{Micro-benchmarks}

\figureref{scaling-microbench} compares the scalability of \nova, webppl, Anglican and Racket for
two microbenchmarks. \verb|fibonacci| implements the naive recursive fibonacci algorithm to test
function call performance. \verb|vectorsum| sums a vector of $n$ ones, to test memory bound
performance.

The results show that \nova, Anglican and Racket all perform similarly for function call
performance, whereas webppl performance is initially similar but does not scale well. For memory
bound performance, webppl performs poorly. The other languages all scale similarly, however \nova is
significantly faster.

\begin{figure}
  \includegraphics[width=0.98\columnwidth]{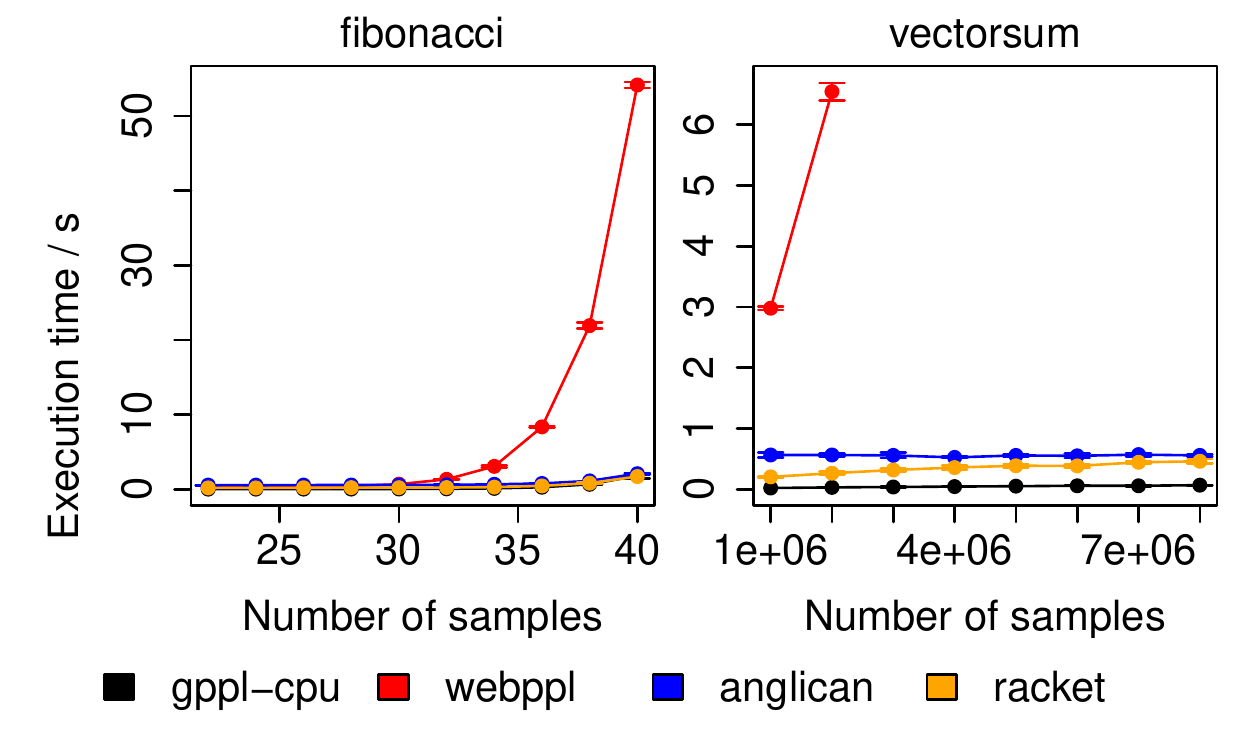}
  \caption{
    Performance results for two micro-benchmarks.
    Error bars show one standard deviation for 10 repeated measurements.
  }
  \figurelabel{scaling-microbench}
\end{figure}

\subsection{Delimited Continuation Performance}
\sectionlabel{delimcont-performance}

The prefix algorithm, shown in \figureref{prefix-code} is used to compare our implementation against
Racket \cite{plt-tr1}. The results are shown in \figureref{prefix-results}, and show that the
performance of our implementation is competitive.

\begin{figure}
  \begin{novacode}
    type List : (+ (Cons : (int, List))
                   (Nil : unit));
    visit <- function (lst) {
      case (lst) {
        Nil = shift(k, Nil())
        Cons (a, rst) =
          (Cons(a,
            shift(k,
              append(k(Nil()),
              reset(k(visit(rst)))))))
      }
    };
    prefix <- function(lst) {
      reset(visit(lst))
    }
  \end{novacode}
  \caption{
    Code for the prefix benchmark in \nova. \code{append} is omitted for brevity.
    For input \code{[1, 2, 3]}, this outputs \code{[1, 1, 2, 1, 2, 3]}.
  }
  \figurelabel{prefix-code}
\end{figure}

\begin{figure}
  \includegraphics[width=0.95\columnwidth]{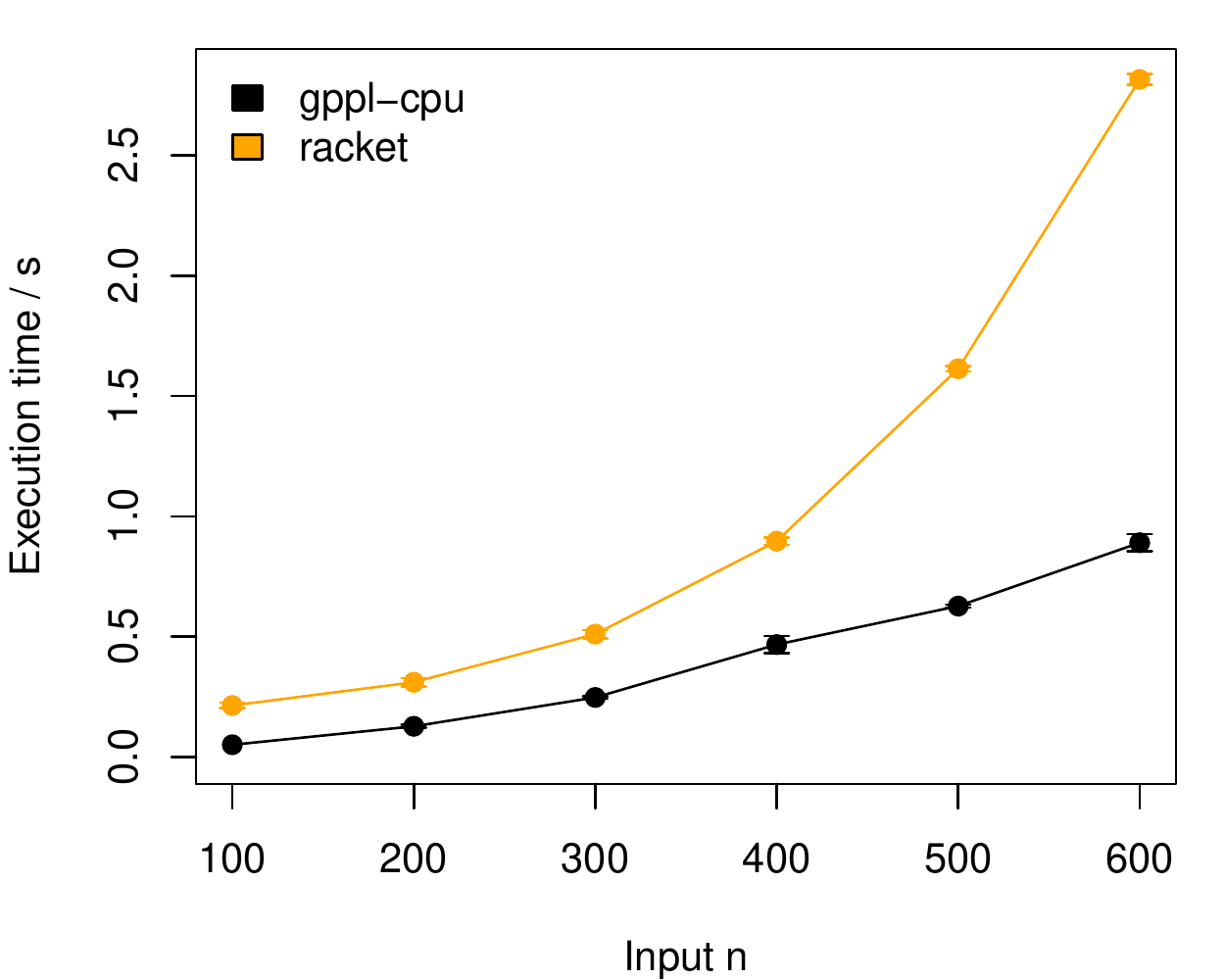}
  \caption{
    Execution time against input size for the prefix benchmark,
    comparing \nova to Racket.
  }
  \figurelabel{prefix-results}
\end{figure}

\section{Conclusions and Future Work}
\sectionlabel{conclusion}

In this paper, we have presented \nova, a probabilistic programming language that generates highly
efficient code for both CPUs and CUDA GPUs. The language is functional in style, and the toolchain
is built on top of LLVM. Our implementation uses delimited continuations on CPU to perform
inference, and custom CUDA codes on GPU.

Compared to other state of the art PPLs Anglican \cite{tolpin2016design} and webppl \cite{dippl},
\nova achieves significantly better performance across a range of benchmarks, and scales linearly to
large numbers of samples. Our implementation of delimited continuations is also competitive with
Racket \cite{plt-tr1}.

In future, we plan to extend our language to support a greater variety of inference methods on both
CPU and GPU. We are investigating parallelizable particle-based inference methods
\cite{paige-nips-2014} for use on GPU.

\bibliography{paper}

\end{document}